\documentclass[12pt]{article}
\usepackage{amsmath}
\usepackage{amsthm}
\usepackage{amssymb}
\usepackage{graphicx}
\usepackage{enumerate}
\usepackage{natbib}
\usepackage{authblk}
\usepackage{float}
\usepackage{amsfonts}
\usepackage{xcolor}
\usepackage{hyperref}
\usepackage[english]{babel}
\newtheorem{theorem}{Theorem}[section]

\newtheorem{assumption}{Assumption}[section]
\usepackage{gensymb}
\usepackage{url} % not crucial - just used below for the URL 

\newcommand{\blind}{1}

\newcommand{\argmin}{\arg\!\min} % AlfC

\begin{document}

\def\spacingset#1{\renewcommand{\baselinestretch}%
{#1}\small\normalsize} \spacingset{1}

%%%%%%%%%%%%%%%%%%%%%%%%%%%%%%%%%%%%%%%%%%%%%%%%%%%%%%%%%%%%%%%%%%%%%%%%%%%%%%

\if1\blind
{
  \title{\bf Reconstructing and Forecasting Marine Dynamic Variable Fields across Space and Time Globally and Gaplessly}
	\author[1$\dag$]{Zhixi Xiong}
	\author[1$\dag$]{Yukang Jiang}
	\author[3]{Wenfang Lu}
	\author[2]{Xueqin Wang}
	\author[1*]{Ting Tian}
	\affil[1]{School of Mathematics, Sun Yat-sen University, Guangzhou, 510275, China} 
	\affil[2]{School of Management, University of Science and Technology of China, Hefei, 230026, China}
	\affil[3]{School of Marine Sciences, Sun Yat-sen University, Guangzhou, 510275, China}
	\affil[*]{Corresponding author. E-mail: tiant55@mail.sysu.edu.cn}
	\affil[ ]{Contributing authors: xiongzhx5@mail2.sysu.edu.cn; jiangyk3@mail2.sysu.edu.cn; luwf6@sysu.edu.cn; wangxq20@ustc.edu.cn;}
	\affil[$\dag$]{These authors contributed equally to this work.}
	\maketitle
} \fi

\if0\blind
{
  \bigskip
  \bigskip
  \bigskip
  \begin{center}
    {\LARGE\bf Reconstructing and forecasting marine dynamic variable fields across space and time globally and gaplessly}
\end{center}
  \medskip
} \fi

%\tableofcontents

\bigskip
\begin{abstract} 
	% 200 words
Spatiotemporal projections in marine science are essential for understanding ocean systems and their impact on Earth's climate. 
However, existing AI-based and statistics-based inversion methods face challenges in leveraging ocean data, generating continuous outputs, and incorporating physical constraints. 
We propose the Marine Dynamic Reconstruction and Forecast Neural Networks (MDRF-Net), which integrates marine physical mechanisms and observed data to reconstruct and forecast continuous ocean temperature-salinity and dynamic fields.
MDRF-Net leverages statistical theories and techniques, incorporating parallel neural network sharing initial layer, two-step training strategy, and ensemble methodology, facilitating in exploring challenging marine areas like the Arctic zone. 
We have theoretically justified the efficacy of our ensemble method and the rationality of it by providing an upper bound on its generalization error.
The effectiveness of MDRF-Net's is validated through a comprehensive simulation study, which highlights its capability to reliably estimate unknown parameters.
Comparison with other inversion methods and reanalysis data are also conducted, and the global test error is 0.455\degree C for temperature and 0.0714psu for salinity.
Overall, MDRF-Net effectively learns the ocean dynamics system using physical mechanisms and statistical insights, contributing to a deeper understanding of marine systems and their impact on the environment and human use of the ocean.
	% 197 words

	% Parameters

\end{abstract} 

\noindent%

\spacingset{1.9} % DON'T change the spacing!
\section{Introduction} 

% Background
Ocean changes quantified by the marine variable fields have significant implications for the economic and social systems that rely on them \citep{tittensor2021next, harley2006impacts}. For example, temperature, salinity, and current behaviors play a crucial role in the ocean state and the performance and sustainability of many organisms \citep{ashton2022predator, smyth2016effects}. {A core challenge in modern AI is the efficiency, exactness, and explainability of analyzing vast, multi-source ocean data, where statistical methods show great promise for extracting valuable insights relevant to both marine science and human activities in this realm. \citep{salcedo2020machine}.} However, the collection of anthropogenically gathered ocean big data from various sources, such as buoys equipped with sensors, satellite remote sensing, and ship-based or airborne radars, has historically been sporadic and limited, resulting in difficulties quantifying continuous marine variability from in situ measurements \citep{johnson2022argo}.  {Improving the comprehension of the oceans' climate impact and facilitating enhanced decision-making across numerous marine-related industries can be achieved by utilizing statistics and AI, which are instrumental in deciphering the immense and varied data available \citep{bauer2015quiet, fox2019challenges}.}
% Ocean-related scientific research and anthropogenic production activities, which are driven by marine variables, will also be impacted, leading to critical consequences for global climate change (\citep{gleckler2012human}). 

% Previous Methods
Recent advancements in predicting the spatial and temporal variability of marine variable fields have demonstrated the potential of AI-based \citep{xiao2019spatiotemporal, xie2019adaptive, song2020novel, su2021super, song2022inversion, zhang2023modified} and statistical inversion methods \citep{lee2019online, yarger2022functional}.  % More details about these research works and background are in Appendix \ref{appendRB}. 
However, AI-based methods often encounter challenges, such as the requirement for time-series gridded data and the inability to provide continuous spatiotemporal predictions. On the other hand, the statistical methods are structurally complex, difficult to implement, and may struggle with handling large-scale data. Furthermore, these methods are purely data-driven, lacking interpretability, and do not fully exploit the relationships between different variable fields. {Employing statistical methodologies and AI techniques plays a critical role in developing an integrated, robust, and interpretable framework for analyzing marine data and predicting ocean states \citep{lou2023application}.}

% Functions of MDRF-Net
To address the aforementioned issues, we propose a mesh-free and easy-to-implement model, Marine Dynamic Reconstruction and Forecast Neural Networks (MDRF-Net), for the reconstruction and forecasting of ocean temperature-salinity and dynamic fields.
{It offers a seamless integration of informative mechanistic equations with diverse data sources and types, grounded in statistical rigor, to achieve both high training efficiency and enhanced accuracy in practical AI applications.}
On one hand, it integrates fundamental physical laws of fluid dynamics, particularly the primitive equations, to ensure predictions are grounded in reality, accounting for temperature-salinity interactions in global fluid movement and mass conservation \citep{hieber2016global, J-LLions_1992}. On the other hand, MDRF-Net adeptly merges relatively granular temperature and salinity data derived from Argo \citep{wong2020argo}, with the meticulous and comprehensive currents reanalysis data provided by \cite{Copernicus2020}.
Thus, MDRF-Net enables the reconstruction and forecasting of complete continuous variable fields based on partial and incomplete observations. 
%Specifically, we have derived seven different variable fields and the uncollected marine variables of density and pressure can be explored and quantified to provide useful environmental evolution. 

% Structure of MDRF-Net
MDRF-Net is built upon the fundamental framework of physics-informed neural networks as developed by Raissi et al. \citep{raissi2019physics}, and three distinct design features of MDRF-Net contribute to its enhanced performance. 
Firstly, employing an ensemble method to rotate the Earth's coordinate system facilitates multiple modeling and weighted averaging of results from diverse sub-learners, allowing for multiple models to work synergistically, compensating for individual weaknesses and improving overall accuracy and robustness of the predictions.
Secondly, employing parallel neural networks that share the initial layer enables the sharing of fundamental information across different variable fields, while the parallel structural design of networks enhances fitting and predictive performance for variable fields in ultra-large-scale data. 
Finally, adopting a two-step training method reduces the redundant calculation of partial derivatives during the initial training stages, thereby easing the computational burden and speeds up the training process without compromising the model's convergence or final performance.

% Theories of MDRF-Net
Theoretically, we show the effectiveness of ensemble method and give an upper bound on the generalization error of MDRF-Net, proving the convergence of the solution provided that the model is adequately trained and the assumptions on the conditional stability estimate of the system of primitive equations hold. At the same time, when the sample domain is significantly smaller than the whole domain, namely, there are a large number of sea areas that cannot be reached by the detectors in reality, other inversion methods without adding mechanisms do not have this convergence, that is, they are unable to give the upper bound of the generalization error.

% Experiments

{MDRF-Net's validation via simulations reveals its superior performance in reconstructing missing data, enhancing field estimation precision, and stably identifying unknown equation parameters compared to similar methods.} It also demonstrates consistent performance with real datasets, adapting to various spatial-temporal prediction tasks and data volumes. Globally, MDRF-Net provides continuous space-time inference of ocean changes and temporal extrapolation, surpassing reanalysis data in certain aspects. It effectively identifies key ocean phenomena, such as the Mediterranean Salinity Crisis and the North Atlantic Warm Current, and monitors hard-to-observe regions like the Arctic.  An interactive R Shiny platform (\url{https://tikitakatikitaka.shinyapps.io/mdrf-net-shiny/}) consolidates variable predictions for user exploration across spaces and times, and the code is accessible at \url{https://github.com/tikitaka0243/mdrf-net}.

% Parametersmm
\section{Methodologies}

\subsection{Primitive equations}\label{sectionPEs}

The goal of our research is to reconstruct and predict the ocean dynamics field, and the primitive equations \citep{hieber2016global, J-LLions_1992}, which describe the ocean dynamics system including current motion and thermohaline diffusion effects, bring important physical information to the table. It is given by Equation \ref{PEs} and contains the momentum (Equation \ref{momentum_1}), hydrostatic balance (Equation \ref{momentum_2}) and continuity (Equation \ref{continuity}) equations, the equations of temperature (Equation \ref{equa_temp}, thermodynamical equation) and salinity (Equation \ref{equa_sal}), and the equation of state (Equation \ref{equa_state}).
\begin{subequations}
	\label{PEs}
	\begin{align}
		\frac{\partial\boldsymbol{v}}{\partial t} + \nabla_{\boldsymbol{v}} \boldsymbol{v} + w\frac{\partial v}{\partial r_a} + \frac{1}{\rho_0} \boldsymbol{\nabla}_h\ p + 
		2\boldsymbol{\omega}_e (\mathbf{e}_r\times\boldsymbol{v})\cos\theta - \zeta\Delta\boldsymbol{v} - \eta\frac{\partial^2\boldsymbol{v}}{\partial r_a^2} & = 0, \label{momentum_1} \\
		\frac{\partial p}{\partial r_a} & = -\rho g, \label{momentum_2} \\
		\text{div}\ \boldsymbol{v} + \frac{\partial w}{\partial r_a} & =0, \label{continuity} \\
		\frac{\partial \tau}{\partial t} + \nabla_{\boldsymbol{v}}\tau + w\frac{\partial\tau}{\partial r_a} - \zeta_\tau\Delta\tau - \eta_\tau\frac{\partial^2\tau}{\partial r_a^2} & = 0, \label{equa_temp} \\
		\frac{\partial \sigma}{\partial t} + \nabla_{\boldsymbol{v}}\sigma + w\frac{\partial\sigma}{\partial r_a} - \zeta_\sigma\Delta\sigma - \eta_\sigma\frac{\partial^2\sigma}{\partial r_a^2} & = 0, \label{equa_sal} \\
		\rho_0 [1 - \beta_\tau(\tau - \tau_0) + \beta_\sigma(\sigma - \sigma_0)] & = \rho. \label{equa_state}
	\end{align}
\end{subequations}

Here we use $\tau, \sigma, w, v_\theta, v_\phi, p,$ and $ \rho$ to denote temperature, salinity, vertical velocity, northward velocity, eastward velocity, pressure, and density respectively, they are all functions of $r, \theta, \varphi, t$, namely radial distance, polar angle (transformed latitude), azimuthal angle (transformed longitude), and time. $\boldsymbol{v} = (v_\theta, v_\varphi)$ represents the horizontal velocity. $r = r_a + r_e$, $r_e$ is the radius of the Earth and $r_a \leq 0$ the vertical coordinate with regard to the sea surface. $\rho_0>0,\ \tau_0>0,\ \sigma_0>0$ are the reference values of the density, temperature and salinity. $\boldsymbol{\omega}_e$ is the vector angular velocity of the Earth and $\mathbf{e}_r$ the vertical unit vector. $\boldsymbol{\nabla}_h$  represents the horizontal gradient operator and $\times$ the cross product operator. $\eta, \zeta$ are eddy viscosity coefficients and $\eta_\tau, \zeta_\tau$ and $\eta_\sigma, \zeta_\sigma$ are eddy diffusivity coefficients for temperature and salinity respectively. Positive expansion coefficients $\beta_\tau,\ \beta_\sigma$, are used, along with the Laplacian $\Delta$, gradient $\nabla_{\boldsymbol{v}}$ with respect to velocity, and divergence operator div.

And there are initial conditions (Equation \ref{ICs}) and boundary conditions (Equation \ref{BCs}), namely space-time boundary conditions:
\begin{equation}
	\label{ICs}
	\boldsymbol{v} = \boldsymbol{i},\ \tau=i_\tau,\ \sigma=i_\sigma,
\end{equation}
\vspace{-30pt}
\begin{subequations}
	\label{BCs}
	\begin{align}
		\frac{\partial\boldsymbol{v}}{\partial r_a} =\boldsymbol{\delta}_{\boldsymbol{v}},\ w=0,\ \frac{\partial\tau}{\partial r_a}+\alpha(\tau-\tau_a)=0,\ \frac{\partial\sigma}{\partial r_a}=0, 
		& \quad \text{on}\ \Gamma_u, \\
		\boldsymbol{v}=\boldsymbol{0},\ w=0,\ \tau=b_\tau,\ \sigma=b_\sigma, 
		& \quad \text{on}\ \Gamma_b, \\
		\boldsymbol{v}=\boldsymbol{0},\ w=0,\ \frac{\partial\tau}{\partial\boldsymbol{\psi}}=0, \ \frac{\partial\sigma}{\partial\boldsymbol{\psi}}=0, 
		& \quad \text{on}\ \Gamma_l, \label{globalBCs}
	\end{align}
\end{subequations}
where $\boldsymbol{i}, i_\tau, i_\sigma$ are the initial values. $\Gamma_u, \Gamma_b$, and $\Gamma_l$ are the upper, bottom, and lateral boundaries of the ocean respectively. The wind stress, denoted as $\boldsymbol{\delta}_{\boldsymbol{v}}$, is contingent upon the velocity of the atmosphere. $\alpha$ is a positive constant associated with the turbulent heating on the ocean's surface, $\tau_a$ is the apparent atmospheric equilibrium temperature, and $b_\tau$ as well as $b_\sigma$, which are functions of the latitude $\theta$ and longitude $\varphi$, represent the temperature and salinity of the sea at the ocean's bottom. On the boundary, the normal vector is $\boldsymbol{\psi}$. More details about the primitive equations are in Appendix A.

%\subsection{Inverse Problem}

	The underlying domain of the primitive equations is $\Omega = [\{r_a\}_{min}+r_e, r_e]\times [0, \pi)\times [0, 2\pi) \in \mathbb{R}$ and the boundary with continuous first order derivatives is $\Gamma = \Gamma_u \cup \Gamma_b \cup \Gamma_l$. We also have the space-time domain and boundary $\bar{\Omega} = \Omega\times [0, T] \subset \mathbb{R}^4$ and $\bar{\Gamma} = \Gamma\times[0, T] \cup \Omega\times\{t=0\}$. Then the primitive equations (Equation \ref{PEs}) can be generally represented as
	\begin{equation}
		\mathcal{F}_{\boldsymbol{\beta}}(\boldsymbol{u})=\boldsymbol{g}
		\label{PEsG}
	\end{equation}
	where $\mathcal{F}: L^2(\bar{\Omega}, \mathbb{R}^6) \mapsto L^2(\bar{\Omega}, \mathbb{R}^6)$ is a general operator for the primitive equations and $L^p(X, Y)$ represents the $p$-norm finite function space mapping from $X$ to $Y$; $\boldsymbol{\beta} = (\beta_\tau, \beta_\sigma)$ represents the unknown parameters of the equations; $\boldsymbol{u} = (\tau, \sigma, w, v_\theta, v_\varphi, p) \in L^2(\bar{\Omega}, \mathbb{R}^6)$ denotes the ocean variable fields and notice that the density field $\rho$ can be derived from temperature field $\tau$ and salinity field $\sigma$; $\boldsymbol{g} \in L^2(\bar{\Omega}, \mathbb{R}^6)$ is the source term and here we only consider the effect of gravity. We also assume that,
	\begin{equation}
		\|\mathcal{F}_{\boldsymbol{\beta}}(\boldsymbol{u})\|_{\mathbb{R}^6} < +\infty,\ \forall\boldsymbol{u}\in\mathbb{R}^6,\ \text{with }\|\boldsymbol{u}\|_{\mathbb{R}^6} < +\infty, \quad \text{and}\quad
		\|\boldsymbol{g}\|_{\mathbb{R}^6} < +\infty.
		\label{H1_H2}
	\end{equation}
	where $\|\cdot\|_{\mathbb{R}^d}$ represents the norm on the $d$-dimensional real space.
	
	The general form of the space-time boundary conditions (Equation \ref{ICs}, \ref{BCs}) is
	\begin{equation}
		\mathcal{B}(\text{tr}(\boldsymbol{u})) = \boldsymbol{b},
		\label{PEsG_boundary}
	\end{equation}
	with the general boundary operator $\mathcal{B}: L^2(\bar{\Gamma}, \mathbb{R}^6) \mapsto L^2(\bar{\Gamma}, \mathbb{R}^6)$, the trace operator $\text{tr}: L^2(\bar{\Omega}, \mathbb{R}^6) \mapsto L^2(\bar{\Gamma}, \mathbb{R}^6)$ and $\boldsymbol{b} \in L^2(\bar{\Gamma}, \mathbb{R}^6)$. All of the operators are bounded under the corresponding norm, that is
	\begin{equation}
		\|\mathcal{B}_{\boldsymbol{\beta}}(\text{tr}(\boldsymbol{u}))\|_{\mathbb{R}^6} < +\infty,\ \forall\boldsymbol{u}\in\mathbb{R}^6,\ \text{with }\|\boldsymbol{u}\|_{\mathbb{R}^6} < +\infty, \quad \text{and}\quad
		\|\boldsymbol{b}\|_{\mathbb{R}^6} < +\infty.
		\label{H3_H4}
	\end{equation}
	It is known that the forward problem of the primitive equations (Equation \ref{PEsG}, \ref{PEsG_boundary}) is well-posed \citep{hieber2016global, charve2008global}, namely, given $\boldsymbol{g} \in L^2(\bar{\Omega}, \mathbb{R}^6)$ and $\boldsymbol{b} \in L^2(\bar{\Gamma}, \mathbb{R}^6)$, there exists a unique solution $\boldsymbol{u}\in L^2(\bar{\Omega}, \mathbb{R}^6)$ satisfying the equations (Equation \ref{PEsG}, \ref{PEsG_boundary}) and continuously depending on changes in boundary conditions.
	
	In ocean dynamic systems, we do not know the complete boundary conditions and the equations' parameters. Thus, the forward problem for the primitive equations will be ill-posed, that is, no unique solution can be guaranteed. However, some observations $\boldsymbol{u}_{\text{obs}}$ of the underlying solution $\boldsymbol{u}$ in a sub-domain $\bar{\Omega}' \subset \bar{\Omega}$ (observation domain) are available, such as temperature and salinity data from Argo \citep{wong2020argo} and three-dimensional velocity data from \cite{Copernicus2020}:
	\begin{equation}
		\boldsymbol{u} = \boldsymbol{u}_{\text{obs}} + \boldsymbol{\epsilon},\ \forall\boldsymbol{x}\in\bar{\Omega},
		\label{PEsG_data}
	\end{equation}
	where $\boldsymbol{\epsilon}$ is a noise term and $\boldsymbol{x} = (r, \theta, \varphi, t)$ represents the space-time coordinate. We assume that
	\begin{equation}
		\|\boldsymbol{u}_{\text{obs}}\|_{L^2(\bar{\Omega}')} < +\infty
		\label{H5}
	\end{equation}
	
	Equations \ref{PEsG}, \ref{PEsG_boundary}, and \ref{PEsG_data} form the inverse problem of the primitive equations, and we assume that it has conditional stability estimate:
	\begin{assumption}[Conditional stability estimate]
		For any $\boldsymbol{u}_1, \boldsymbol{u}_2 \in L^2(\bar{\Omega}, \mathbb{R}^6)$, the primitive  equations saisfy
		\begin{equation}
			\begin{aligned}
				\|\boldsymbol{u}_1-\boldsymbol{u}_2\|_{L^2(S)} \leq\ & C\left(\|\boldsymbol{u}_1\|_{L^2(\bar{\Omega})}, \|\boldsymbol{u}_2\|_{L^2(\bar{\Omega})}\right) \cdot  \left( 
				\|\boldsymbol{u}_1 - \boldsymbol{u}_2\|^{\gamma'}_{L^2(\bar{\Omega}')}  
				+
				\right. \\ & \left.
				\|\mathcal{F}_{\boldsymbol{\beta}}(\boldsymbol{u}_1) - \mathcal{F}_{\boldsymbol{\beta}}(\boldsymbol{u}_2)\|^{\gamma_{1}}_{L^2(\bar{\Omega})} +
				\|\mathcal{B}(\text{tr}(\boldsymbol{u}_1)) - \mathcal{B}(\text{tr}(\boldsymbol{u}_2))\|^{\gamma_{2}}_{L^2(\bar{\Gamma})}
				\right)
			\end{aligned}
			\label{conditional_stability}
		\end{equation}
		for some $0< \gamma',\gamma_1, \gamma_2\leq 1$ and for any subset $\bar{\Omega}'\subset S\subset\bar{\Omega}$.
	\end{assumption}

\subsection{Marine Dynamic Reconstruction and Forecast Neural Networks (MDRF-Net)} 
\label{SECpenn}

In order to enable the reconstruct and forecast ocean dynamic fields, including temperature, salinity, vertical, northward and eastward velocities, and pressure fields, we constructed the Marine Dynamic Reconstruction and Forecast Neural Networks (MDRF-Net), which seamlessly integrates information from the data and the primitive equations in a concise structure by embedding the equations into the loss function. And since Copernicus' current reanalysis data is already highly refined and comprehensive, our focus is on using it as a bridge to connect the primitive equations and enabling MDRF-Net to reconstruct the temperature and salinity fields with more accurate and comprehensive information about the ocean current system.

%Our work starts with the inversion of the ocean temperature and salinity fields using the groundbreaking Argo dataset, combined with the Copernicus current reanalysis field which was initialized at the beginning of 2020. The Copernicus data serves as a bridge, connecting the primitive equations and enabling MDRF-Net to reconstruct the temperature and salinity fields with more accurate and comprehensive information about the ocean current system. The basic physical principles of these variables are described by the primitive equations (Section \ref{sectionPEs}). As a result, MDRF-Net seamlessly integrates information from the data and the primitive equations in a concise structure by embedding the equations into the loss function of the neural networks.
	
\begin{figure}[H]
	\centering
	\includegraphics[width=\textwidth]{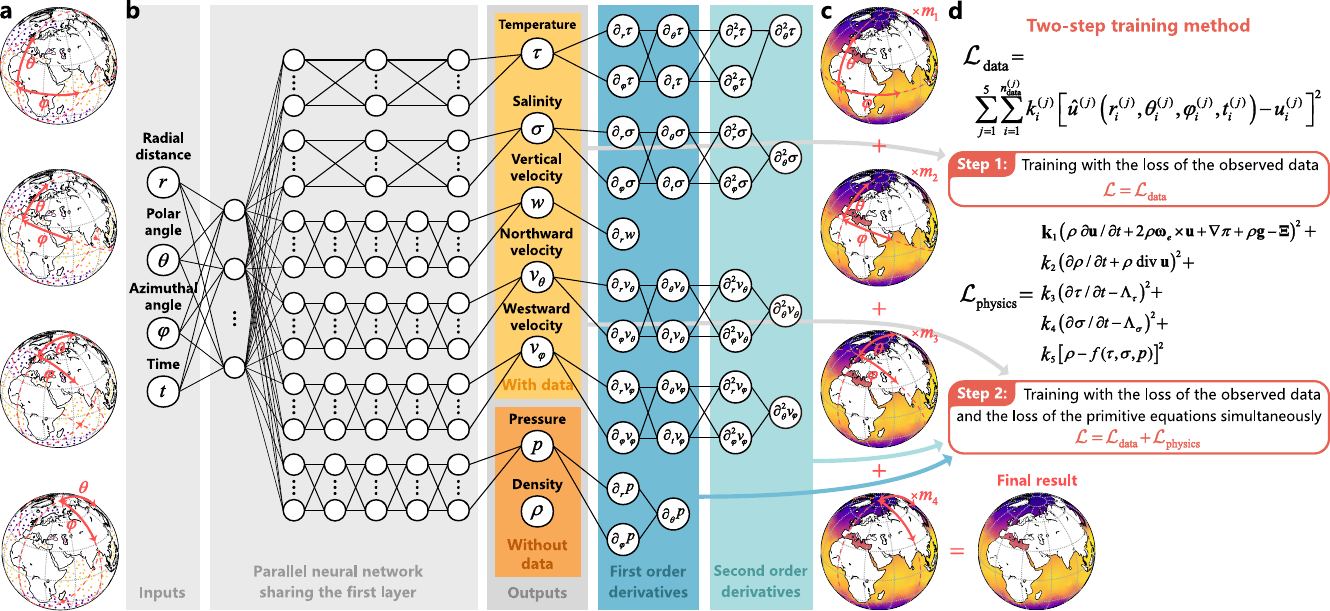}
	
	\caption{\linespread{1.2}\selectfont\textbf{Marine Dynamic Reconstruction and Forecast Neural Networks (MDRF-Net)}. Before the training data is fed into the neural network, its coordinate system is rotated multiple times along the 0-degree meridian (\textbf{a}). The main body of MDRF-Net is a parallel neural network that shares the first layer (\textbf{b}), with inputs of coordinates and time, and outputs the variables that need to be differentiated in the primitive equation. There are three layers of the fully connected neural networks for the temperature and salinity fields, and five layers for the other four fields. The first and second-order partial derivatives of each ocean variable are calculated to compute the loss of equations. Finally, the results derived in each coordinate system are weighted and averaged to obtain the final outcome (\textbf{c}). In the two-step training strategy (\textbf{d}), the parameters of neural networks are optimized based on the loss of data in the first step and on the loss of data and physics together in the second step. The formula for $\mathcal{L}_{\text{physics}}$ here is a simplified version, see the Materials and Methods section for the precise implementation.}
	\label{PENN_structure}
\end{figure}

Inside MDRF-Net (Figure \ref{PENN_structure}), there is a parallel fully connected neural network that shares the first layer. This design allows for the adaptation of the neural networks to different sizes of marine variable fields with varying complexities. 
Additionally, we propose a two-step training strategy where parameters are first optimized using observed data, and then using the loss of physics and the observed data simultaneously. The second step adds the primitive equations into the loss function to provide spatial-temporal physics information. This approach eliminates the need for the model to calculate a large number of partial derivatives when its outputs deviate from expectations, resulting in faster and more efficient training. 
To improve the model's performance at high latitudes, we also use an ensemble method by rotating the Earth's coordinate system multiple times along the 0° longitude (prime meridian), thus modeling the model multiple times and taking a weighted average of the results.
% Overall, MDRF-Net offers greater flexibility in integrating different types of data and mechanism equations, with high training efficiency and prediction accuracy. 
More detailed information about MDRF-Net can be found in the Appendix C. 

\subsection{Implementation of MDRF-Net}\label{sectionMDRF-Net}

Consider an input $\boldsymbol{x}\in\bar{\Omega}$, we represent the neural network of MDRF-Net by the following:
\begin{equation}
	\begin{aligned}
		\boldsymbol{u}^{(j)}_\Theta(\boldsymbol{x}) = & \ C^{(j)}_{K^{(j)}}\circ \sigma_{\tanh} \circ C^{(j)}_{K^{(j)}-1} \circ \cdots\circ \sigma_{\tanh} \circ C^{(j)}_{2} \circ \sigma_{\tanh} \circ C_{1}(\boldsymbol{x}), \\
		\hat{\boldsymbol{u}}(\boldsymbol{x}) = & \left[\boldsymbol{u}^{(1)}_\Theta(\boldsymbol{x}), \boldsymbol{u}^{(2)}_\Theta(\boldsymbol{x}), \cdots, \boldsymbol{u}^{(6)}_\Theta(\boldsymbol{x})\right].
	\end{aligned}
\end{equation}
where $\boldsymbol{u}_\Theta(\boldsymbol{x})$ is the variable fields represented by a parallel neural network sharing the first layer with parameters $\Theta$; $\boldsymbol{u}^{(j)}_\Theta(\boldsymbol{x})$ is the $j^{\text{th}}$ sub-network for the $j^{\text{th}}$ variable field; $\circ$ is the composition of the functions and $\sigma_{\tanh}$ is the tanh activation function. $C^{(j)}_k$ represents the linear transformation of the $k^{\text{th}}$ layer of the $j^{\text{th}}$ sub-network, and
\begin{equation}
	C^{(j)}_{k}(\boldsymbol{x}^{(j)}_k) = \boldsymbol{W}^{(j)}_{k}\boldsymbol{x}^{(j)}_{k} + \boldsymbol{b}^{(j)}_{k},\ 1\leq j\leq6,\ 2\leq k\leq K^{(j)},
\end{equation}
where $\boldsymbol{W}^{(j)}_{k} \in \mathbb{R}^{d^{(j)}_{k+1}\times d^{(j)}_{k}}$ is the weights matrix, $\boldsymbol{b}^{(j)}_{k}\in \mathbb{R}^{d^{(j)}_{k+1}}$ is the bias term and $\boldsymbol{x}^{(j)}_{k}\in \mathbb{R}^{d^{(j)}_{k}}$ is the output of the $(k-1)^{\text{th}}$ layer. Here, we set $K^{(1)} = K^{(2)} = 3$ and $K^{(3)} = K^{(4)} = \cdots = K^{(6)} = 5$; $d^{(j)}_2 = d^{(j)}_3 = \cdots = d^{(j)}_{K^{(j)}} = 128,\ d^{(j)}_{K^{(j)} + 1} = 1,\ 1\leq j\leq6$. For the first shared layer, similarly, $C_1(\boldsymbol{x}_1) = \boldsymbol{W}_1\boldsymbol{x}_1 + \boldsymbol{b}_1,\ \boldsymbol{W}_1\in\mathbb{R}^{d_2^{(j)}\times4},\ \boldsymbol{x}_1\in\mathbb{R}^4,\ \boldsymbol{b}_1\in\mathbb{R}^{d_2^{(j)}},\ 1\leq j\leq6.$
Then the parameters of the neural network are $\Theta = \{\boldsymbol{W}_k^{(j)}, \boldsymbol{b}_k^{(j)}, \boldsymbol{W}_1, \boldsymbol{b}_1\},\ \forall 1\leq k\leq K^{(j)},\ 1\leq j\leq6$.

Given the primitive equations with initial and boundary conditions and the observed data, the residuals of the data and physics information for MDRF-Net $\boldsymbol{u}_\Theta$ and the equations' unknown parameters $\boldsymbol{\beta}$ are $\mathcal{L}_{\text{data}}(\boldsymbol{u}_\Theta):= \boldsymbol{u}_\Theta - \boldsymbol{u}_{\text{obs}} - \boldsymbol{\epsilon},$
$\mathcal{L}_{\text{pde}}(\boldsymbol{u}_\Theta, \boldsymbol{\beta}):= \mathcal{F}_{\boldsymbol{\beta}}(\boldsymbol{u}_\Theta) - \boldsymbol{g}, \ 
	\text{and}\ 
	\mathcal{L}_{\text{icbc}}(\boldsymbol{u}_\Theta):= \mathcal{B}(\text{tr}(\boldsymbol{u}_\Theta)) - \boldsymbol{b},$
By the first three assumptions (Equation \ref{H1_H2}, \ref{H3_H4}, \ref{H5}), we know that $\mathcal{L}_{\text{data}}\in L^2(\bar{\Omega}', \mathbb{R}^6),\ \mathcal{L}_{\text{pde}}\in L^2(\bar{\Omega}, \mathbb{R}^6),\ 
\mathcal{L}_{\text{icbc}}\in L^2(\bar{\Gamma}, \mathbb{R}^6),$ and
$\|\mathcal{L}_{\text{data}}\|_{L^2(\bar{\Omega}')}$, $\|\mathcal{L}_{\text{pde}}\|_{L^2(\bar{\Omega})}$ $\|\mathcal{L}_{\text{icbc}}\|_{L^2(\bar{\Omega})} < +\infty$, for all $\Theta\in \boldsymbol{\Theta},\ \boldsymbol{\beta} \in B$, $\boldsymbol{\Theta}$ and $B$ are the parameter spaces of the neural network and the primitive equations respectively.

Therefore, the optimization problem of MDRF-Net is
\begin{equation}
	\Theta^*, \boldsymbol{\beta}^* = \argmin_{\substack{\Theta\in\boldsymbol{\Theta}\\ \boldsymbol{\beta} \in B}} \left\{
	\left\| \mathcal{L}_{\text{data}} \right\|_{L^2(\bar{\Omega}')} 
	+  
	\lambda'_1\left\| 	\mathcal{L}_{\text{pde}} \right\|_{L^2(\bar{\Omega})} 
	+  
	\lambda'_2\left\| 	\mathcal{L}_{\text{icbc}} \right\|_{L^2(\bar{\Gamma})} 
	\right\}.
\end{equation}
Equivalently,
\begin{equation}
	\Theta^*, \boldsymbol{\beta}^* = \argmin_{\substack{\Theta\in\boldsymbol{\Theta}\\ \boldsymbol{\beta} \in B}} \left\{  
	\int_{\bar{\Omega}'} \left| \mathcal{L}_{\text{data}} \right|^2 d\boldsymbol{x}
	+  
	\lambda_1\int_{\bar{\Omega}} \left| \mathcal{L}_{\text{pde}} \right|^2 d\boldsymbol{x}
	+  
	\lambda_2\int_{\bar{\Gamma}} \left| \mathcal{L}_{\text{icbc}} \right|^2 d\boldsymbol{x}
	\right\}.
\end{equation}
where $\lambda'_1, \lambda_1$ and $\lambda'_2, \lambda_2$ are additional regularization hyper-parameters, and the physics residual terms can be considered as penalization terms. Additionally, the problem is able to be approximated using the quadratic rules, with the coordinates of the observations $\{\boldsymbol{x}'_i\}$ with all $1\leq i\leq N_{\text{data}}$, combined with the selected sampling points inside the space-time domain $\{\boldsymbol{x}_{1,i}\},\ \boldsymbol{x}_{1,i}\in \bar{\Omega},\ 1\leq i\leq N_{\text{pde}}$ and points at the space-time boundary $\{\boldsymbol{x}_{2,i}\},\ \boldsymbol{x}_{2,i} \in \bar{\Gamma},\ 1\leq i\leq N_{\text{icbc}}$. Then we have the following loss function:
\begin{equation}
	J(\Theta, \boldsymbol{\beta}) := \sum_{i=1}^{N_{\text{data}}} k'_i \left| \mathcal{L}_{\text{data}} (\boldsymbol{x}'_i) \right|^2 + \lambda_1 \sum_{i=1}^{N_{\text{pde}}} k_{1,i} \left| \mathcal{L}_{\text{pde}} (\boldsymbol{x}_{1,i}) \right|^2
	+ \lambda_2 \sum_{i=1}^{N_{\text{icbc}}} k_{2,i} \left| \mathcal{L}_{\text{icbc}} (\boldsymbol{x}_{2,i}) \right|^2
\end{equation}
where $k'_i$, $k_{1,i}$ and $k_{2,i}$ are weights. 
When calculating the errors of the observed data ($\mathcal{L}_{\text{data}}$), the sampling points are naturally the positions in space and time of the samples in the training data set $(r^{(j)}_i, \theta^{(j)}_i, \varphi^{(j)}_i, t^{(j)}_i), i = 1, 2,\ldots, N^{(j)}_{\text{data}}, \ j = 1, 2, \ldots, 5$; and when computing the loss of the primitive equations ($\mathcal{L}_{\text{pde}}$), the sampling points can come from the location of the observed data, or can be generated additionally within the equations' underlying domain $\bar{\Omega} = \Omega\times[0, T]$. In this study, additional gridded sampling points are generated for calculating the equations' loss in the global scenario due to the non-uniform distribution of observed data locations. It is important to note that the number of observations strictly at the initial time or definition domain boundaries is usually small. Therefore, separate sampling points need to be generated for calculating the errors in the initial conditions and boundary conditions ($\mathcal{L}_{\text{icbc}}$) based on the definition domain. 
% However, in the local scenario, the sampling points used are randomly generated within the definition domain and boundary since the area is a simple cuboid.

During the training of MDRF-Net, the partial derivatives of first and second order of the outputs $\boldsymbol{u}_\Theta$ with respect to the inputs $\boldsymbol{x}$ are cleverly computed by the automatic differentiation algorithm required in the optimization process of neural networks. As in classical neural networks, MDRF-Net uses the gradient descent method based on the back propagation algorithm to optimize the parameters. For the unknown parameters of the primitive equations, MDRF-Net optimizes them together with the parameters of the neural networks, thus solving the `inverse problem' of the primitive equations.

Regarding the ensemble method, suppose we have inversion results for both the original and rotated variable fields, represented as $\hat{\boldsymbol{u}}^{(r)}, r=1,\ldots,N_{ro}$. Then the final weighted result at the polar angle $\theta$ is
\begin{equation}
	\hat{\boldsymbol{u}}(\theta) = \frac{\sum_{r=1}^{N_{ro}} m_r(\theta_r) \hat{\boldsymbol{u}}^{(r)}(\theta_r)}{\sum_{r=1}^{N_{ro}} m_r(\theta_r)} ,
\end{equation}
where $m_r(\theta_r) = \text{logistic}(10(\theta_r / \pi - 0.5))$, $\theta_r$ is the polar angle of the rotated coordinates.

\subsection{Estimation of Generalization Error}\label{sectionGE}

The generalization error is the error of MDRF-Net on unseen data. We set $\bar{\Omega}' \subset S \subset \bar{\Omega}$ and define the corresponding generalization error as
\begin{equation}
	\mathcal{E}(S) = \mathcal{E}\left(
	S; \Theta^*, \boldsymbol{\beta}^*,
	\left\{\boldsymbol{x}_i'\right\}_{i=1}^{N_{\text{data}}},
	\left\{\boldsymbol{x}_{1,i}\right\}_{i=1}^{N_{\text{pde}}},
	\left\{\boldsymbol{x}_{2,i}\right\}_{i=1}^{N_{\text{icbc}}}
	\right) = \|\boldsymbol{u} - \hat{\boldsymbol{u}}\|_{L^2(S)},
\end{equation}
which depends on the train sampling points, data and the optimized MDRF-Net's parameters $\Theta^*$ and equations' unknown parameters $\boldsymbol{\beta}^*$; $\hat{\boldsymbol{u}} = \boldsymbol{u}_{\Theta^*}$ represents the trained MDRF-Net with $\Theta^*$ as its parameters. 

Due to neural networks' inability to well capture the variation pattern of longitudinal density with latitude, we design an ensemble method involving multiple rotations in spherical coordinates, and its effectiveness can be given by the following theorem.

\begin{theorem}[The effectiveness of ensemble method]
	Let $\boldsymbol{u} \in L^2(\bar{\Omega}, \mathbb{R}^6)$ be the solution of the inverse
	problem of the primitive equations, and $\hat{\boldsymbol{u}}$, $\hat{\boldsymbol{u}}^{(r)}$ are the trained MDRF-Net and the $r^{\text{th}}$ sub-learner. Then the generalization error fo MDRF-Net with ensemble method satisfies
	\begin{equation}
		\|\boldsymbol{u} - \hat{\boldsymbol{u}}\| \leq \| \boldsymbol{u} - \hat{\boldsymbol{u}}^{(r)} \|, \quad \forall 1\leq r\leq N_{ro}
	\end{equation}
\end{theorem}

We can estimate the generalization error in terms of the training error $\mathcal{E}_{\text{train}} = \mathcal{E}_{\text{train}}(\Theta^*,
\{\boldsymbol{x}_i'\}_{i=1}^{N_{\text{data}}}$,
$\{\boldsymbol{x}_{1,i}\}_{i=1}^{N_{\text{pde}}},
\{\boldsymbol{x}_{2,i}\}_{i=1}^{N_{\text{icbc}}})$, defined by:
\begin{equation}
	\mathcal{E}_{\text{train}} = \left(
	\mathcal{E}_{\text{data}}^2 
	+ \lambda_1 \mathcal{E}_{\text{pde}}^2
	+ \lambda_2 \mathcal{E}_{\text{icbc}}^2
	\right)^{\frac{1}{2}},
\end{equation}
where $\mathcal{E}_{\text{data}} := (\sum_{i=1}^{N_{\text{data}}} k'_i \left| \mathcal{L}_{\text{data}} (\boldsymbol{x}'_i) \right|^2)^{\frac{1}{2}},$ 
$\mathcal{E}_{\text{pde}} := (\sum_{i=1}^{N_{\text{pde}}} k_{1,i} \left| \mathcal{L}_{\text{pde}} (\boldsymbol{x}_{1,i}) \right|^2)^{\frac{1}{2}},\ 
	\text{and}\ 
	\mathcal{E}_{\text{icbc}} := (\sum_{i=1}^{N_{\text{icbc}}} k_{2,i} \left| \mathcal{L}_{\text{icbc}} (\boldsymbol{x}_{2,i}) \right|^2)^{\frac{1}{2}}.$
The bound on generalization error in terms of training error $\mathcal{E}_{\text{train}}$ is given by the following estimate \citep{10.1093/imanum/drab032}:

\begin{theorem}[Upper bound of generalization error]
	Let $\boldsymbol{u} \in L^2(\bar{\Omega}, \mathbb{R}^6)$ be the solution of the inverse
	problem of the primitive equations. Assume the conditional	stability estimate assumption (Equation \ref{conditional_stability}) holds for any	$S$, s.t. $\bar{\Omega}' \subset S\subset\bar{\Omega}$. Let $\hat{\boldsymbol{u}}$ be the trained MDRF-Net, based on the training sampling points $\{\boldsymbol{x}_{1,i}\}_{i=1}^{N_{\text{pde}}},\ \{\boldsymbol{x}_{2,i}\}_{i=1}^{N_{\text{icbc}}}$ and the observed data coordinates $\{\boldsymbol{x}'_i\}_{i=1}^{N_{\text{data}}}$. Additionally, assume the residuals $\mathcal{L}_{\text{pde}}$, $\mathcal{L}_{\text{icbc}}$ and $\mathcal{L}_{\text{data}}$ be such that $\mathcal{L}_{\text{data}} \in L^2(\bar{\Omega}', \mathbb{R}^6)$, $\mathcal{L}_{\text{pde}} \in L^2(\bar{\Omega}, \mathbb{R}^6)$ and $\mathcal{L}_{\text{icbc}} \in L^2(\bar{\Gamma}, \mathbb{R}^6)$ and the quadrature error is bounded. Then the following estimate on the generalization error holds:
	\begin{equation}
		\begin{aligned}
			\mathcal{E}(S) \leq\ & C\left(
			\mathcal{E}^{\gamma'}_{\text{data}} +
			\mathcal{E}^{\gamma_1}_{\text{pde}} + \mathcal{E}^{\gamma_2}_{\text{icbc}} + 
			\|\boldsymbol{\epsilon}\|^{\gamma'}_{L^2(\bar{\Omega}')} + \right. \\
			& \left.
			\left(C'_q\right)^{\frac{\gamma'}{2}}
			N^{-\frac{a'\gamma'}{2}}_{\text{data}} +
			\left(C_{1,q}\right)^{\frac{\gamma_1}{2}}
			N_{\text{pde}}^{-\frac{a_1\gamma_1}{2}} +
			\left(C_{2,q}\right)^{\frac{\gamma_2}{2}}
			N_{\text{icbc}}^{-\frac{a_2\gamma_2}{2}}
			\right),
		\end{aligned}
	\end{equation}
	with constants $C = C(\|\boldsymbol{u}\|_{L^2(\bar{\Omega})}, \|\hat{\boldsymbol{u}}\|_{L^2(\bar{\Omega})})$, 
			$C'_q = C'_q(\left\||\mathcal{L}_{\text{data}}|^2\right\|
			_{L^2(\bar{\Omega}')})$,
			$C_{1,q} = C_q(\left\||\mathcal{L}_{\text{pde}}|^2\right\|
			_{L^2(\bar{\Omega})})$,
			$C_{2,q} = C_q(\left\||\mathcal{L}_{\text{icbc}}|^2\right\|
			_{L^2(\bar{\Gamma})}).$
	\label{theorem_bound}
\end{theorem}

Notice that Theorem \ref{theorem_bound} is satisfied by the precondition that the conditional stability estimate (Equation \ref{conditional_stability}) of the primitive equations hold. And when MDRF-Net has no constraints from the ocean dynamics mechanism, that is, the primitive equations are not embedded in the loss function, the model degenerates into a purely data-driven neural network. At this point we can view the training of the model as a special `inverse problem', i.e., $\mathcal{F}_{\boldsymbol{\beta}} (\boldsymbol{u}) = c_1 \ \text{and}\   \mathcal{B}(\boldsymbol{u}) = c_2 \  \text{for any }\boldsymbol{u} \in L^2(\bar{\Omega}, \mathbb{R}^6),$
where $c_1$ and $c_2$ are constants and the observations (Equation \ref{PEsG_data}) remains the same. Now assume that the measure of the integration domain of the observation space $\bar{\Omega}'$ is $\varrho_i$ times of the whole space $\bar{\Omega}$, $0<\varrho_i < 1$ for all $1\leq i \leq 4$. Then the norm 
\begin{equation}
	\begin{aligned}
		\|\boldsymbol{u}_1 - \boldsymbol{u}_2\|^{\gamma'}_{L^2(\bar{\Omega}')} 
		& =
		\left(
		\int_{\bar{\Omega}'} \left|\boldsymbol{u}_1 - \boldsymbol{u}_2\right|^2 d\boldsymbol{x} 
		\right) ^{\frac{\gamma'}{2}} \\
		& = \left(
		\prod_{i=1}^{4} \varrho_i
		\int_{\bar{\Omega}} 
		\left|\boldsymbol{u}_1 - \boldsymbol{u}_2\right|^2 d\boldsymbol{x} 
		\right) ^{\frac{\gamma'}{2}}
		= \left(\prod_{i=1}^{4} \varrho_i\right)^{\frac{\gamma'}{2}}
		\|\boldsymbol{u}_1 - \boldsymbol{u}_2\|^{\gamma'}_{L^2(\bar{\Omega})},
	\end{aligned}
\end{equation}
which can not be control by the constant $C\left(\|\boldsymbol{u}_1\|_{L^2(\bar{\Omega})}, \|\boldsymbol{u}_2\|_{L^2(\bar{\Omega})}\right)$, since $\varrho_i$ is determined by the observation space $\bar{\Omega}'$. Hence, there exists a space $S'$ satisfying $\bar{\Omega}'\subset S'\subset\bar{\Omega}$ that
\begin{equation}
	\begin{aligned}
		\|\boldsymbol{u}_1-\boldsymbol{u}_2\|_{L^2(S')} >\ & C\left(\|\boldsymbol{u}_1\|_{L^2(\bar{\Omega})}, \|\boldsymbol{u}_2\|_{L^2(\bar{\Omega})}\right) \cdot
		\|\boldsymbol{u}_1 - \boldsymbol{u}_2\|^{\gamma'}_{L^2(\bar{\Omega}')}  
	\end{aligned}
	\label{conditional_stability_not_satisfied}
\end{equation}
for any $0< \gamma'\leq 1$. Then the conditional stability estimate cannot be satisfied and an upper bound on the generalization error cannot be given. 
In fact, this derivation applies to all methods that do not incorporate mechanism, such as Gaussian Process Regression and Regression Kriging, which will be discussed later. This implies that when the data domain is not sufficiently large, these methods lack an upper bound on their generalization error.

\section{Results}

\subsection{Simulation Study}

We first explore the capabilities of MDRF-Net in a simulated system by considering a 2D simplified version (Equation \ref{PEs_2d}) of the primitive equations (Equation \ref{PEs}) which has only one dimension in the horizontal direction and does not include the diffusion equation for salinity (Equation \ref{equa_sal}) as well as the equation of state (Equation \ref{equa_state}), so it has only four variables temperature $\tau$, horizontal velocity $v$, vertical velocity $w$ and pressure $p$. Unlike Equation \ref{PEs}, the underlying domain of this primitive equations is based on a Cartesian coordinate system rather than a spherical coordinate system, and it is dimensionless, that is, all variables take on values without units.
\begin{equation}
	\begin{aligned}
		\frac{\partial v}{\partial t} + v \frac{\partial v}{\partial x} + w \frac{\partial v}{\partial z} - \eta \frac{\partial^2 v}{\partial x^2} - \zeta \frac{\partial^2 v}{\partial z^2} + \frac{\partial p}{\partial x} & =0, \\
		\frac{\partial p}{\partial z} & =-\tau, \\
		\frac{\partial v}{\partial x} + \frac{\partial w}{\partial z} & =0, \\
		\frac{\partial \tau}{\partial t} + v \frac{\partial \tau}{\partial x} + w \frac{\partial \tau}{\partial z} - \eta_\tau \frac{\partial^2 \tau}{\partial x^2} - \zeta_\tau \frac{\partial^2 \tau}{\partial z^2} & = Q,
	\end{aligned}
	\label{PEs_2d}
\end{equation}
This system of equations has a specific Taylor-Green vortex solution for a specific periodic source term $Q$ \citep{hu2023higherorder},
\begin{equation}
	\begin{aligned}
		v & = -\sin (2\pi x)\cos(2\pi z)\exp \left[-4\pi^2(\eta+ \zeta)t \right],\\
		w & = \cos(2\pi x)\sin(2\pi z)\exp\left[-4\pi^2 (\eta + \zeta) t\right], \\
		p & = \frac{1}{4} \cos (4\pi x)\exp \left[ -8\pi^2(\eta + \zeta)t\right] + \frac{1}{2\pi} \cos (2\pi z)\exp(-4\pi^2\zeta_\tau t),\\
		\tau & = \sin (2\pi z)\exp (-4\pi^2 \zeta_\tau t), \\
		Q & = \pi \cos (2\pi x)\sin(4\pi z)\exp \left[-4\pi^2 (\eta + \zeta + \zeta_\tau)t\right].
	\end{aligned}
\end{equation}

Accordingly, we set $\eta=\zeta=0.01,\ \zeta_\tau=0.02$, and generated 1000 samples randomly in the data domain and without pressure variables to mimic the reality of the ocean data. Note that the Taylor-Green vortex is independent of the value of $\zeta_\tau$. In reconstructing these variable fields using MDRF-Net, we set the parameters $\zeta$ and $\zeta_\tau$ to be unknown with initial value 0 and the value of $\eta$ is correlated with $\zeta$, which would be unrecognizable if they were both set to be unknown. We also examine the performance of MDRF-Net without mechanism (N-MDRF-Net), and the commonly used marine variable field interpolation methods Gaussian Process Regression (GPR) and Regression Kriging (R-Kriging), which support ungridded data and provide continuous inversion results, on simulated datasets.

\begin{figure}[H]
	\centering
	\includegraphics[width=\textwidth]{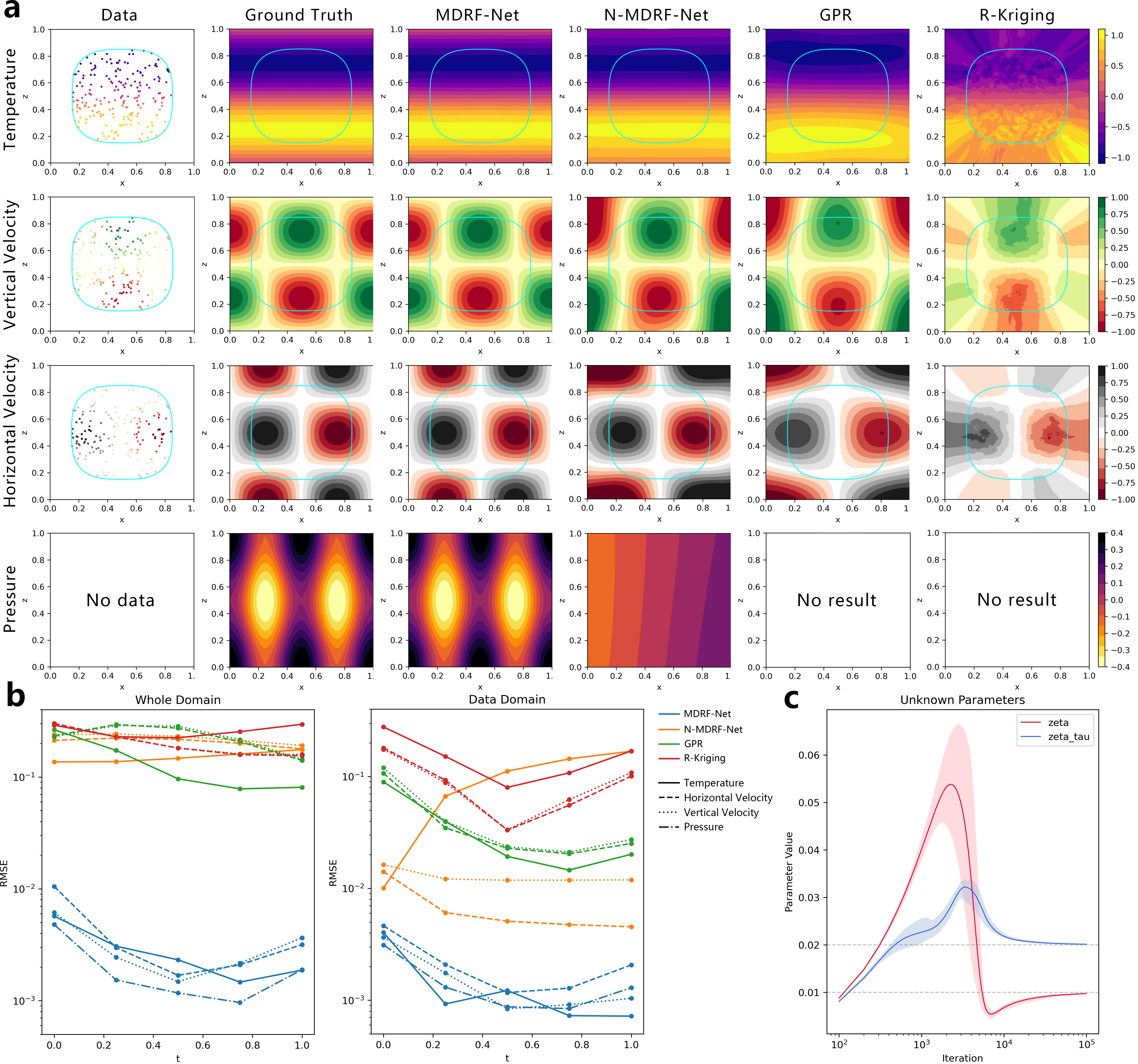}
	\caption{\linespread{1.2}\selectfont\textbf{Simulation study.} MDRF-Net, MDRF-Net without mechanism (N-MDRF-Net) are compared with Gaussian Process Regression (GPR) and Regression Kriging (R-Kriging) at $t$=0 (\textbf{a}, more results can be found in Appendix E). Comparison of the accuracy of competing methods over time in the whole domain and the data domain, the accuracies of the pressure fields except for MDRF-Net are not plotted (\textbf{b}). The data are generated within rounded rectangles and do not contain the pressure variables. Changes of the unknown parameters during training, the solid lines are the means of the results of 100 repetitions of the experiment, while the light-colored areas in the background are the point-wise 95\% confidence intervals (\textbf{c}).} 
	\label{simulation_comparison}
\end{figure}

MDRF-Net perfectly reconstructs the real variable fields, even in domain where there is no data and for the variable without data. The other models can only reconstruct the variable fields fairly well in the domain with data, while it is basically a failure in the domain without data. For pressure fields with no data at all, GPR and R-Kriging fail to give results completely, whereas N-MDRF-Net still gives an output because it uses a parallel neural network that shares the first layer (Figure \ref{simulation_comparison} \textbf{a}).

In terms of accuracy, the overall root mean square error (RMSE) of MDRF-Net is below $10^{-2}$, which is far superior to the other methods over the whole domain, even in domain where data is available, which is consistent with the expectations derived from theoretical analyses. Notice that even though R-Kriging is not as continuous as the N-MDRF-Net and GPR results, the three methods that do not include the mechanism are at the same level of overall RMSE. In the data domain, N-MDRF-Net slightly outperforms GPR and R-Kriging, however, its predicted temperature fields exhibit a tendency to accumulate substantially larger errors over extended temporal horizons. In addition, the RMSEs of all the models show more or less a decreasing and then increasing pattern throughout the time span (Figure \ref{simulation_comparison} \textbf{b}).

During the training process of MDRF-Net, the two unknown parameters, $\zeta$ and $\zeta_\tau$, steadily approach their true values of 0.01 and 0.02 from below and above, respectively. Notably, in the early stages of the solution procedure, considerable fluctuations are observed in the values of these unknowns, with $\zeta$, for instance, momentarily reaching around 0.06 (Figure \ref{simulation_comparison} \textbf{c}).

\subsection{Assessment of MDRF-Net with Real Data} 
\label{Performance_Evaluation}
We assess the performance of MDRF-Net in the equatorial Pacific (Appendix B) and analyze error trends over space and time (Figures \ref{MC} \textbf{a,b,c}; additional errors in Appendix F). These trends prove steady, suggesting MDRF-Net effectively captures spatial and temporal patterns. Errors near islands are marginally higher but show a distinct peak at depths of 250-350m for all variables except vertical velocity, which follows a separate pattern (see Appendix F). When tested against Argo and Copernicus data, MDRF-Net yields satisfactory RMSEs: 0.358°C for temperature, 0.0474 psu for salinity, and for velocities-1.955$\times 10^{-5}$ m/s (vertical), 0.0465 m/s (northward), and 0.0513 m/s (eastward), demonstrating its competence in predicting ocean conditions.

 MDRF-Net is compared with the same competing models as the simulation study. We compute the RMSEs of all models utilizing training sets of different sizes on the test sets of the original spatiotemporal domain, two extended spatial domains (outwardly stretching and inwardly filling), and an extended temporal domain (forecasting). MDRF-Net excels in all evaluations, particularly in 3-month forecasting (Figure \ref{MC} \textbf{d iv}), outperforming traditional statistical inversing methods. It significantly outmatches GPR and R-Kriging for temperature and salinity, and dominates in flow velocity predictions, especially with larger datasets. As data volume increases, MDRF-Net's errors decrease sharply. Its incorporation of primitive equations further boosts accuracy in inferring unobserved pressure and density fields, highlighting the equations' critical role in augmenting the model's performance with additional physics-driven insights.

\begin{figure}[H]
	\centering
	\includegraphics[width=0.85\textwidth]{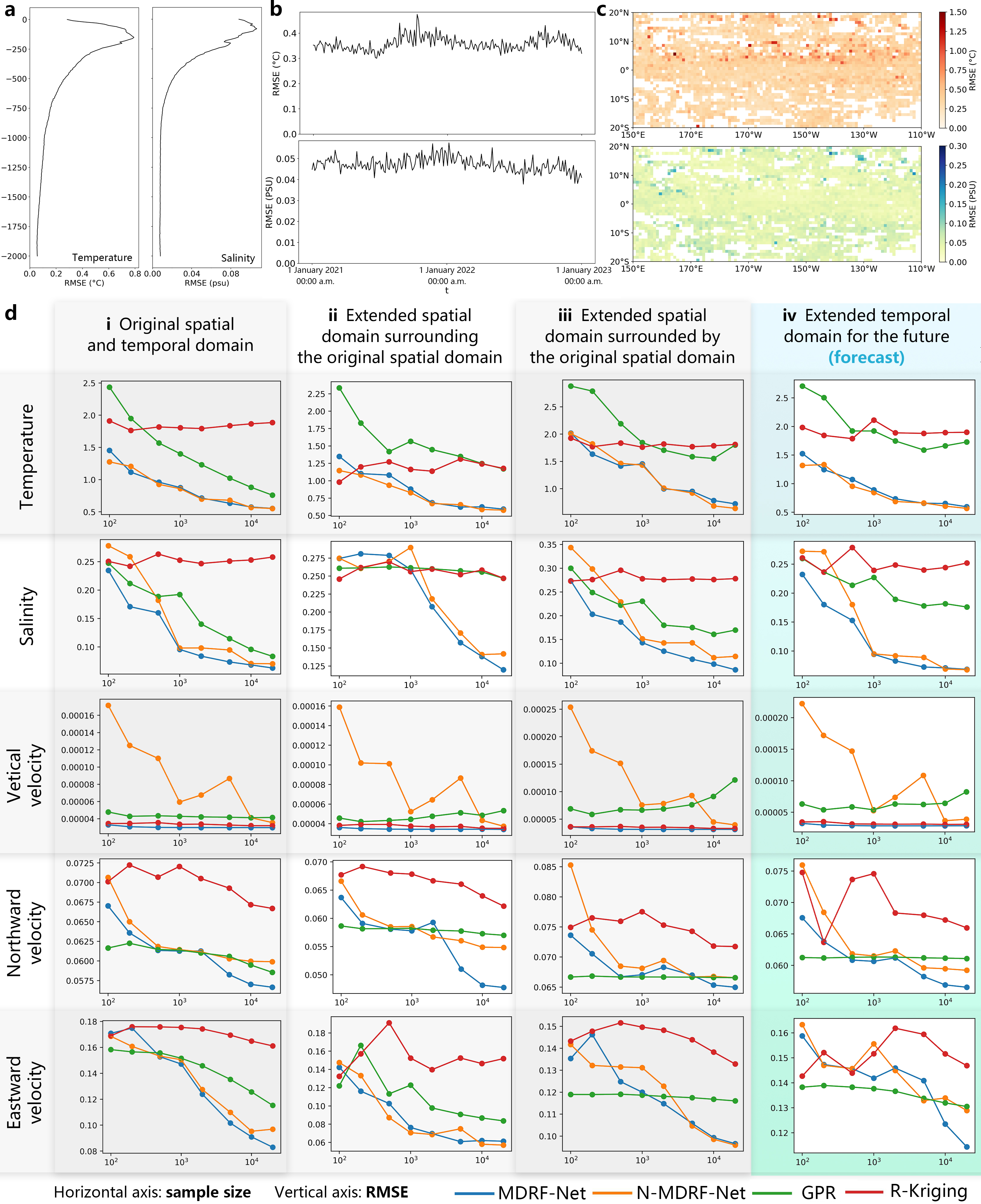}
	\caption{\linespread{1.2}\selectfont\textbf{The trends of root-mean-square errors (RMSEs) with the test set.} The variability of RMSEs is shown over depth (\textbf{a}), time (\textbf{b}), and coordinates (\textbf{c}) for temperature and salinity. The blanks in (\textbf{c}) represent islands. The RMSE charts for all marine variables over depth, time and coordinates can be found in Appendix F. Comparisons (\textbf{d}) are made to evaluate the performance of MDRF-Net in relation to its version that does not include the primitive equations (N-MDRF-Net), as well as the Gaussian process regression (GPR) and Regression Kriging algorithms (R-Kriging).  Panel (\textbf{i}) represents the inversion accuracy of the models on the original spatiotemporal scale, while panels (\textbf{ii}), (\textbf{iii}) depict the prediction effects on the extended spatial scales (outwardly stretching and inwardly filling), respectively. Panel (\textbf{iv}) shows the 3-month average forecasting errors. scale All error values presented are averaged over five replicated experiments.}
	\label{MC}
\end{figure}

\subsection{Reconstructing Global Oceanic Variations over Space Continuously} 
\label{SECinversionoutcomes}

%   \subsubsection*{Local scenario}
%     \label{Inversion_Results}

% \subsubsection*{Global scenario}

In order to characterize global marine activity, we have expanded the application of MDRF-Net to the global ocean, and local scenario of equatorial Pacific is presented in Appendix G. To enhance MDRF-Net's performance along the coastlines, we have implemented Neumann boundary conditions for temperature and salinity, along with Dirichlet boundary conditions for current flow (Equation \ref{globalBCs}). Despite slightly lower accuracy compared to smaller sea areas, MDRF-Net delivers impressive results, even enabling the inversion of two variable fields without collected data. The overall RMSE values for temperature, salinity, vertical velocity, northward velocity, and eastward velocity fields are 0.455 °C, 0.0714 psu, 4.254$\times 10^{-6}$ m/s, 0.0777 m/s, and 0.0825 m/s respectively.

\begin{figure}[H]
\begin{center}
	\includegraphics[width=0.88\textwidth]{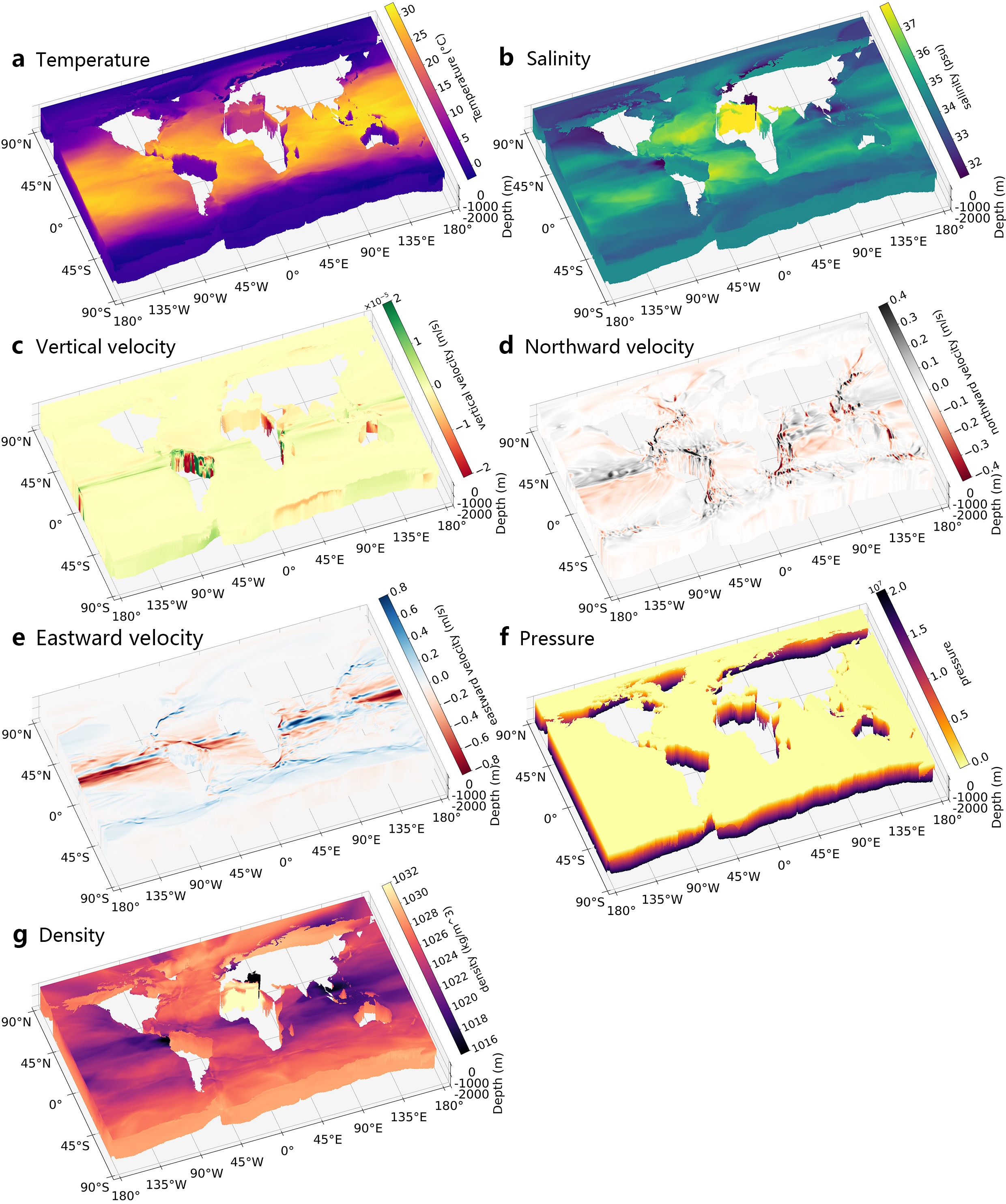}
\end{center}
	\caption{\linespread{1.2}\selectfont\textbf{Global inversion performances of temperature (\textbf{a}), salinity (\textbf{b}), vertical velocity (\textbf{c}), northward velocity (\textbf{d}), eastward velocity (\textbf{e}), pressure (\textbf{f}) and density (\textbf{g}) fields at 16 January 2021, 12:00 a.m.}  The pressure (\textbf{f}) and density (\textbf{g}) fields are inverted without observed data. Results for more time spots can be found in Appendix G, while the full performances for 2021 and 2022 can be found in the R Shiny platform (https://tikitakatikitaka.shinyapps.io/mdrf-net-shiny/). \label{fit_WW_summary}}
\end{figure}

The inversion results for the five variable fields with available observations show good performance. 
The temperature, salinity, eastward velocity, and northward velocity fields exhibit clear patterns (Figure \ref{fit_WW_summary} \textbf{a,b,d,e}).
For example, the temperature field decreases with increasing latitude at the surface and remains relatively constant at deeper levels (Figure \ref{fit_WW_summary} \textbf{a}).
The salinity field shows extreme values in specific regions such as the Mediterranean Sea and the Black Sea, as well as subsurface occurrences in the Arctic Ocean (Figure \ref{fit_WW_summary} \textbf{b}). The three current fields display higher currents near the equator and changes corresponding to land formations (Figure \ref{fit_WW_summary} \textbf{c,d,e}).
For instance, the northward currents (Figure \ref{fit_WW_summary} \textbf{d}) are stronger on the east side of the continent than on the west side. Regarding the pressure and density fields (Figure \ref{fit_WW_summary} \textbf{f,g}) without available observed values, they exhibit expected patterns consistent with local scenarios. The pressure field correlates primarily with water depth, increasing with greater depths. The density field, on the other hand, shows a maximum and minimum in a region similar to that of the salinity field.

%\begin{figure}
%\begin{center}
%\includegraphics[width=3in]{fig1.pdf}
%\end{center}
%\caption{Consistency comparison in fitting surrogate model in the tidal
%	power example. \label{fig:first}}
%\end{figure}

\begin{figure}[H]
	\centering 
	\includegraphics[width=0.87\textwidth]{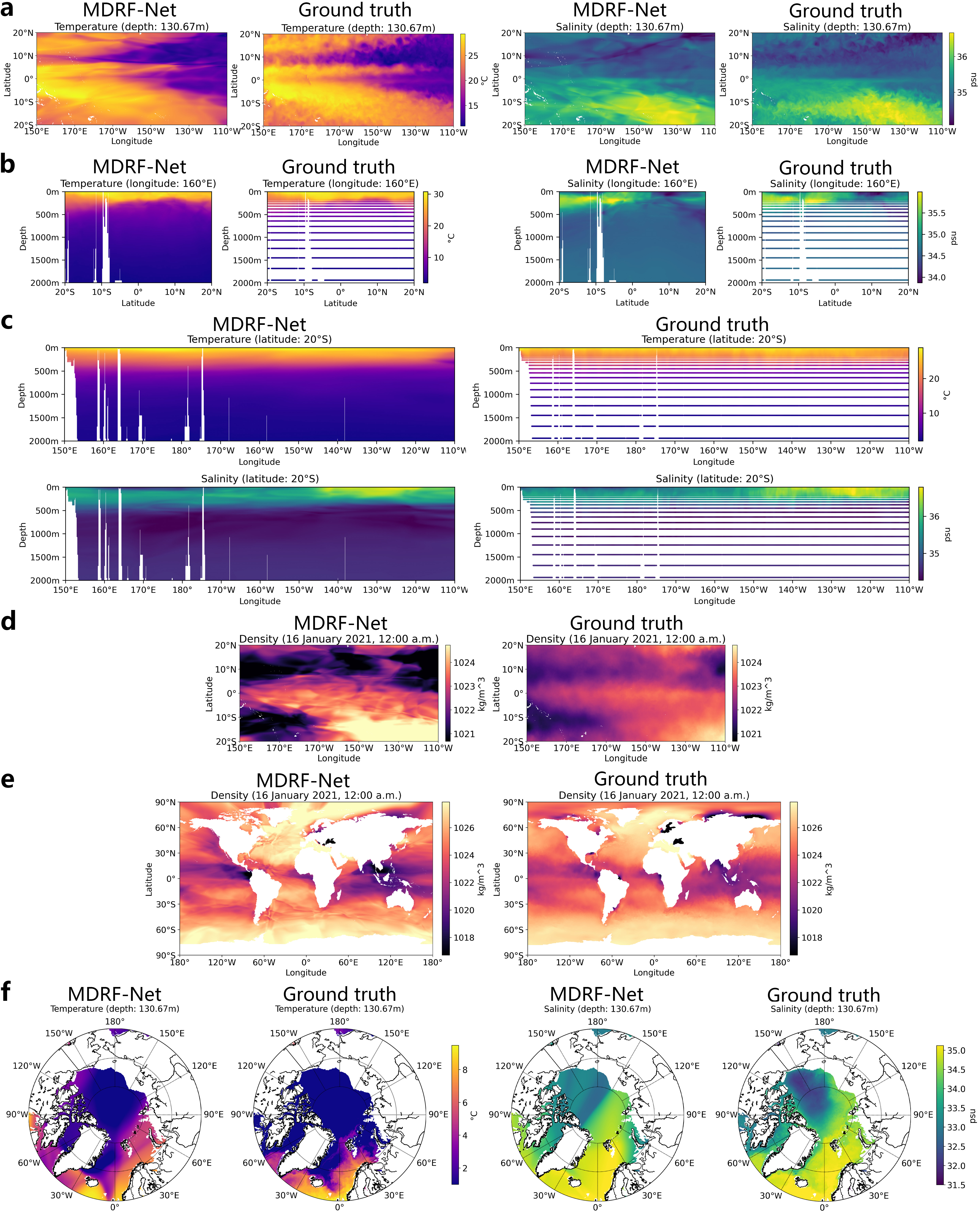} 
	\caption{\linespread{1.2}\selectfont\textbf{The comparison between the inversion performances of the temperature, salinity and density fields of MDRF-Net and the reanalyzed data from Copernicus, labeled as ground truth.} Comparisons are conducted at the depth (\textbf{a}), longitude (\textbf{b}), and latitude cross-sections (\textbf{c}). The blank areas in the images represent islands. Note that the layering seen in the Copernicus images is a result of the dataset's sparse coverage of depth. Within the range of 2000 meters, only 40 layers are available for the reanalyzed data and there is only sea surface density data available. Panels (\textbf{d}) and (\textbf{e}) depict the density fields for the local and global scenarios respectively. The performance of MDRF-Net for temperature and salinity fields at the Arctic zone are shown (\textbf{f}). The more comparison can be found in Appendix G.}
	\label{fit_comparison1}
\end{figure}

From the direct comparison with the reanalyzed fields, it is evident that MDRF-Net is able to invert temperature and salinity fields that closely resemble their reanalyzed counterparts, despite not being directly trained using the reanalyzed data (Figure \ref{fit_comparison1} \textbf{a,b,c}). This is achieved through the backfeeding of the exact current fields using the bridging of the primitive equations. Additionally, our variable fields demonstrate continuity in both space and time. For instance, in Figure \ref{fit_comparison1} \textbf{b,c}, the reanalysis field shows a discontinuity in water depth and clear stratification. It should be noted that Copernicus' reanalysis fields only include 40 layers up to a depth of 2000 meters.

The comparison of the density field illustrates the greater advantage of MDRF-Net (Figure \ref{fit_comparison1} \textbf{d}), as it is not based on collected data like the pressure field. Instead, MDRF-Net inverts the density field using observations of other variable fields and the original equations exclusively, particularly through the equation of state. While we do not have an actual representation of the pressure field, and even the density field in the Copernicus reanalysis dataset only contains data from the ocean surface, MDRF-Net is capable of inverting the density field for the three-dimensional(3D) ocean and shows similarity to the reanalysis field at the ocean surface. The density field shows better results than the local scenario in the surface density versus the reanalyzed field (Figure \ref{fit_comparison1} \textbf{e}), with a very elegant match of patterns. Furthermore, we showcase the inversion performances of the temperature and salinity fields of MDRF-Net at the Arctic zone (Figure \ref{fit_comparison1} \textbf{f}), and there are still some differences in the Arctic Ocean in Northern Canada, but when compared to the reanalysis data, the pattern of the MDRF-Net inversion results is roughly the same.

\subsection{Forecasting Global Oceanic Variations over Time Gaplessly} 
% Adding new results here by Zhixi

\begin{figure}[H]
	\centering 
	\includegraphics[width=0.6\textwidth]{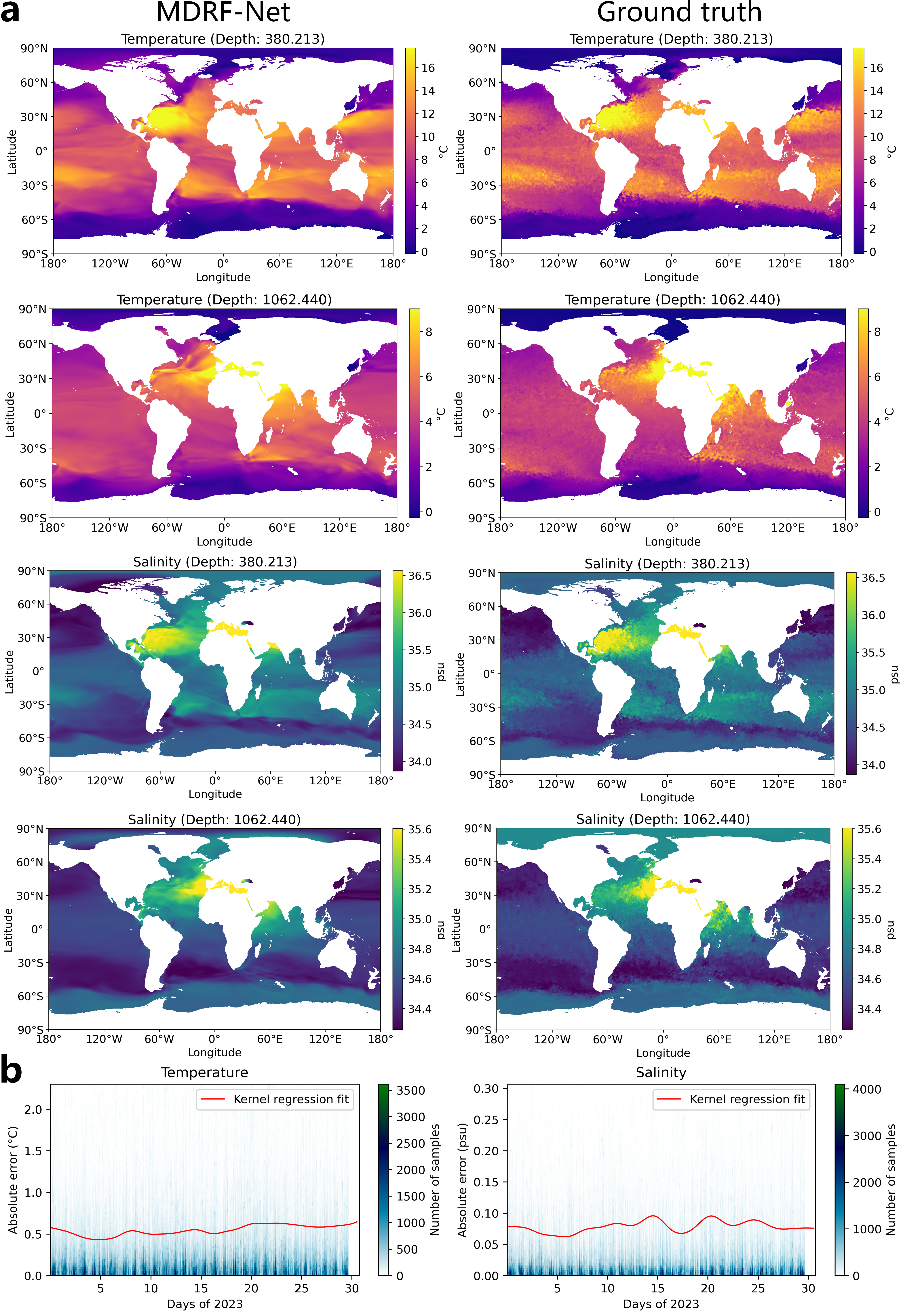} 
	\caption{\linespread{1.2}\selectfont\textbf{Long-term forecasting.} 30 days forecasting (30 January 2023, 12:00 a.m.) results of temperature and salinity are compared with the corresponding reanalysis data (\textbf{a}). More forecast time spans are shown in Appendix G. The absolute errors of 1-month forecasting are calculated with the Argo data (\textbf{b}). The red line is the kernel regression result of 5000 samples from all the data, with bandwidth 1.}
	\label{forecasting}
\end{figure}

In addition to projecting variable fields onto a gapless space and time scale, MDRF-Net can deliver uninterrupted forecasts. We give short- (1 and 7 days) and long-term (30 days) projections of MDRF-Net and compare them against reanalysis data for the same time-frame. In the short-term predictions (Appendix G), the MDRF-Net results show a high degree of agreement with the reanalysis variable fields, especially the temperature and salinity fields, while the 3D current field appears smoother than the reanalysis field, but the main current patterns are still well recognized by MDRF-Net. For long-term forecasting (Figure \ref{forecasting} \textbf{a}), it can be observed that most forecast results closely align with the reanalysis data across different depths with minor discrepancies observed in some land-marginal seas like the Black Sea and the Red Sea. Notably, MDRF-Net accurately captures features such as a warmer zone in the western Atlantic Ocean at a depth of 380.213 meters and the spread of high salinity from the Mediterranean Sea into the Atlantic Ocean at a depth of 1,062.440 meters.

Quantitatively, MDRF-Net demonstrates remarkable accuracy and consistent stability in forecasting, evident from the two-dimensional distributions of absolute errors (Figure \ref{forecasting} \textbf{b}). These visualizations highlight that a large portion of prediction errors in the test set cluster closely around zero. Furthermore, in trend analysis, MDRF-Net displays exceptional consistency. Over a one-month period, there is a slight observable upward trend, but this minor deviation does not detract from its overall robust and steady performance that remains consistent throughout the evaluation period.

\section{Discussion} 
% Novelty and Significance of MDRF-Net
{MDRF-Net addresses a core AI problem in contemporary oceanography by integrating neural networks with ocean dynamics' fundamental laws. With its ensemble method, shared-layer full connectivity, and two-step training process, MDRF-Net contributes to statistical methodology, theory, and application, demonstrating the potential of statistics in empowering AI tools to bridge data-driven models with physical oceanography.  This innovative approach employs statistical theories and methodologies to offer unique insights into marine phenomena, advancing real-world scientific understanding.}

% Integration of Multiple Data Sources
MDRF-Net excels in integrating multiple datasets seamlessly, fusing Argo observations with Copernicus reanalysis data for a robust inversion base. It infers seven oceanic variables from five observed ones, deepening our insights. Its design promotes easy data source integration, adapting to varied sample sizes and ranges. Utilizing 2+ million Argo profiles and hundreds of millions of reanalysis points, MDRF-Net's multi-source data fusion has the potential to significantly improve the generalization capability and accuracy of forecasts in large-scale ocean \citep{guillou2020wave, adhikary2023improved}.

% Efficiency and Extensibility %%%
MDRF-Net stands out for its remarkable extensibility, adeptly managing variable marine data at scale. It inverts two uncollected fields (density, pressure) from five known variables and extends this capability to poorly monitored areas like the Arctic and deep sea \citep{ura2013observation, charrassin2008southern}. By facilitating future change predictions, MDRF-Net supports proactive decision-making and resource management \citep{ellefmo2023marine}, with its swift forecasting prowess making it a powerful tool for anticipating ocean dynamics.

% Estimation of generalization error
MDRF-Net's generalization error estimation reveals its effectiveness in inversely solving the primitive equation for ocean variables, meeting three key criteria: a well-trained model, indicated by low training error; appropriate regularization enabled by the smooth tanh activation, ensuring accurate residual approximation via quadrature; fulfillment of the conditional stability assumption for the inverse problem. Consequently, MDRF-Net successfully approximates the inverse problem and yields precise reconstructions and forecasts of ocean dynamics. As for those methods that do not include physics mechanisms (N-MDRF-Net, GPR, R-Kriging), there are no upper bounds on their generalization errors, and their theoretical convergence is not guaranteed. We also show that the ensemble method of rotating the earth coordinates yields lower generalisation errors and improves the model's performance in polar regions (Appendix Figure 5).

% Simulation Study
Simulation studies indicate that MDRF-Net excels in reconstructing both partial and complete variable fields with missing data, consistently delivering higher accuracy than other methods, even in data-rich domains. {This aligns with the theoretical results, specifically, in regions outside the rounded rectangular area where data is absent (Figure \ref{simulation_comparison}), other methods lacking mechanistic insights fail to make effective predictions, thereby resulting in unbounded generalization errors in those locales.} Meanwhile, MDRF-Net is stable and recognizable for the unknown parameter positional parameters in the dynamical system of the simulation experiment.

% Comparison with Existing Methods
MDRF-Net distinguishes itself from both AI-based \citep{xiao2019spatiotemporal, xie2019adaptive, song2020novel, su2021super, song2022inversion, zhang2023modified} and statistical inversion models \citep{lee2019online, yarger2022functional} by effectively handling diverse sampling locations and offering continuous results. Its adaptability to sparse spatiotemporal data, coupled with natural interpretability and structural simplicity, facilitates seamless data integration, achieving higher accuracy and efficiency. MDRF-Net thus emerges as a superior solution, precisely capturing marine dynamics compared to other existing methodologies (Figure \ref{MC} \textbf{d}).

% Global and Continuous Predictions
Few methods can efficiently integrate and invert crucial variables across the global ocean in a continuous spatiotemporal manner. Upon comparing our MDRF-Net results with reanalyzed data (Figure \ref{fit_comparison1}), it is evident that MDRF-Net's consistently superior performance in global scenarios positions it as a formidable tool for conducting comprehensive marine studies. The continuous provision of global inversion marine fields allows for a nuanced understanding of long-term trends and patterns, offering a unique advantage in the field. MDRF-Net excels in making reasonable inferences about ocean changes that exhibit continuity both spatially and temporally.

% Forecasting %%%
Forecasting over extended periods poses a major challenge, especially for global, high-resolution demands requiring substantial computational efficiency \citep{wolff2020statistical, o2015characterizing}. MDRF-Net addresses this by delivering high-precision forecasts for multiple climate variables, excelling in both short- and long-term projections. Comparative analysis with reanalysis data reveals MDRF-Net's exceptional performance in extrapolating forecasts over time. Notably, MDRF-Net exhibits remarkable accuracy in temperature and salinity fields, with extrapolation errors only slightly higher than those of interpolation. Ocean velocity forecasts benefit from incorporated mechanics, yielding smoother outputs than the turbulent reanalysis data (Appendix G). Crucially, MDRF-Net's gapless predictions for various variables across the globe and time span do not heavily tax computational resources, advancing four-dimensional oceanic forecasting with enhanced resolution compared to current high-resolution forecasting methods \citep{wolff2020statistical, zhang2023preliminary}.

% Applications and Implications
MDRF-Net has broad real-world implications, uncovering previously unseen patterns. Its reconstruction of 4D oceanic fields enhances understanding and guides marine activities, vital for studying impacts on marine life, nutrients, and environment stability \citep{behrenfeld1997photosynthetic, anderson2021marine, sarmiento2006ocean}. By elucidating events like the Mediterranean Salinity Crisis \citep{hsu1977history, ryan2009decoding} and the North Atlantic Warm Current \citep{schmitz1993north, jiang2021multicentennial, gunnarsson2021recent}, MDRF-Net showcases historical and climatic insight. Its practical applications span fisheries management \citep{carnevale2022marine}, maritime navigation \citep{melia2016sea, kelly2020they}, and offshore infrastructure resilience \citep{el2022experimental}, while also finely monitoring tropical phenomena such as El Ni\~{n}o \citep{cai2021changing}, highlighting versatility and practical significance.

% Conclusion
{MDRF-Net's mesh-free methodology addresses a pivotal challenge in marine scientific AI, amplifying its effectiveness through statistical insights. This model, which addresses limitations in existing methods and difficulties of variable field inversion, showcases the unique role of statistical theories in improving AI methods. }
While external forces' impact on shallow coastal areas is acknowledged for future refinement, MDRF-Net has demonstrated excellent overall performance and generalization capabilities.
{MDRF-Net's innovative application in sustainable marine resource management, underlines the significant contribution of statistical methodologies to AI, offering solutions for real-world science advancement.}

\section*{Data and Code Availability} 
We select worldwide temperature and salinity data collected by the lifting floats from the Argo program (\url{https://argo.ucsd.edu/}), and eastward, northward and vertical reanalysis seawater velocity data from the EU Copernicus Marine Service (\url{https://marine.copernicus.eu/}) for the two years 2021 to 2022 and water depths up to 2000 m. 

The code for this article has been uploaded to Github at the link (\url{https://github.com/tikitaka0243/mdrf-net}). We utilized the Python library DeepXDE \citep{lu2021deepxde} to construct the MDRF-Net.

%\subsection{Argo and Copernicus Data Sets}\label{SECdata}

% 	\begin{figure}[H]
	% 	\centering     %%% not \center
	% 	\subfigure[Argo buoys distribution up to April 18, 2023]{
		% 		\label{argo_distribution}
		% 		\includegraphics[height=30mm]{image/argo_status}
		% 	}
	% 	\subfigure[Argo buoys working cycle]{
		% 		\label{argo_cycle}
		% 		\includegraphics[height=30mm]{image/float_cycle_1}
		% 	}
	% 	\caption{Information about the Argo project}
	% 	\label{argo_info}
	% \end{figure}

% Taking into account the amount of data and the nature of the ocean physical system, in the local scenario, 

%Due to restrictions on float deployment, the Argo data is relatively sparse and uneven, while the reanalysis data covered the entire horizontal area of the ocean, but it is still relatively discrete in the vertical direction.
%Overall, we use the entire Argo dataset with more than $10^8$ size of temperature and salinity observations and sample equal amounts of current data from the Copernicus reanalysis field.
%More details about the data are in Appendix B.

% Temperature and salinity data collected by the lifting buoys can be found on the Argo program website (\url{https://argo.ucsd.edu/}), while reanalysis data of currents and density are provided by the EU Copernicus Marine Service (\url{https://marine.copernicus.eu/}). They also provide ocean topography data with a fairly high degree of accuracy.

\if1\blind
{

\section*{Acknowledgments} 
This research was supported by the National Key R\&D Program of China (No. 2022YFA1003800), the National Natural Science Foundation of China  (No. 71991474; No. 12001554; No. 72171216), the Key Research and Development Program of Guangdong, China (No. 2019B020228001), the Science and Technology Program of Guangzhou, China (No. 202201011578), and the Natural Science Foundation of Guangdong Province, China (No. 2021A1515010205). The funding agencies had no role in the study design, data collection, analysis, decision to publish, or preparation of the manuscript.
} \fi

\section*{Disclosure Statement} 
The authors report there are no competing interests to declare.

%\begin{figure}
%\begin{center}
%\includegraphics[width=3in]{fig1.pdf}
%\end{center}
%\caption{Consistency comparison in fitting surrogate model in the tidal
%power example. \label{fig:first}}
%\end{figure}

%\begin{table}
%\caption{D-optimality values for design $X$ under five different scenarios.  \label{tab:tabone}}
%\begin{center}
%\begin{tabular}{rrrrr}
%one & two & three & four & five\\\hline
%1.23 & 3.45 & 5.00 & 1.21 & 3.41 \\
%1.23 & 3.45 & 5.00 & 1.21 & 3.42 \\
%1.23 & 3.45 & 5.00 & 1.21 & 3.43 \\
%\end{tabular}
%\end{center}
%\end{table}

\bigskip
\begin{center}
{\large\bf SUPPLEMENTARY MATERIAL}
\end{center}

\begin{description}

\item[Title:] Supplementary Material for  ``Reconstructing and Forecasting Marine Dynamic Variable Fields across Space and Time Globally and Gaplessly"

\item[Overview:] The derivation of the primitive equations is provided in \textbf{Appendix A}. More details about the Argo and Copernicus data sets are provided in \textbf{Appendix B}. More details about the Marine Dynamic Reconstruction and Forecast neural network (MDRF-Net) are provided in \textbf{Appendix C}. More derivations of Generalization Error Estimation are provided in \textbf{Appendix D}. More results of simulation study are provided in \textbf{Appendix \textsc{E}}. The error analysis of MDRF-Net is provided in \textbf{Appendix F}. Additional reconstruction and forecast results are provided in \textbf{Appendix G}.

\end{description}

\linespread{1.2}\selectfont
\bibliographystyle{Chicago}

\bibliography{Bibliography-MM-MC}
\end{document}